\input mssymb.tex  

 \def\CC{\hbox{{$\cal C$}}}
 
\def\CL{\hbox{{$\cal L$}}}

\def\CR{\hbox{{$\cal R$}}}

\def\C{{\Bbb C}}
\def\Z{{\Bbb Z}}

\def\vecu{{\bf u}}

\def\<{\langle}
\def\>{\rangle}
\def\del{{\partial}}
\def\lform{\hbox{$\sqcup$}\llap{\hbox{$\sqcap$}}}

\def\eps{{\epsilon}}

\def\cosub{{\Delta\kern -.65em \raisebox{.02em}{-}\kern .35em}}

\def\tens{\mathop{\otimes}}
\def\la{{\triangleright}}
\def\isom{{\cong}}

\def\Aut{{\rm Aut}}

\def\Ad{{\rm Ad}}

\def\id{{\rm id}}

\def\proof{\goodbreak\noindent{\bf Proof\quad}}

\def\endproof{{\ $\lform$}\bigskip }

\def\und#1{{\underline {#1}}}

\def\o{{}_{\scriptscriptstyle(1)}}
\def\t{{}_{\scriptscriptstyle(2)}}
\def\th{{}_{\scriptscriptstyle(3)}}

\def\note#1{}

\def\equad{\kern -1.7em}
\def\qqquad{\qquad\quad}

\def\eqn#1#2{\begin{equation}#2\label{#1}\end{equation}}
\def\cmath#1{\[\begin{array}{c} #1 \end{array}\]}

\def\ceqn#1#2{\begin{equation}\label{#1}
\begin{array}{c}#2\end{array}\end{equation}}
\def\align#1{\begin{eqnarray*}#1\end{eqnarray*}}

\documentstyle[11pt, epsf]{article}
\newtheorem{lemma}{Lemma}[section]
\newtheorem{propos}[lemma]{Proposition}
\newtheorem{example}[lemma]{Example}
\newtheorem{theorem}[lemma]{Theorem}

\newtheorem{corol}[lemma]{Corollary}

\begin{document}\baselineskip 22pt

{\ }\hskip 4.2in DAMTP/93-64
\begin{center} {\Large SOLUTIONS OF THE YANG-BAXTER EQUATIONS FROM BRAIDED-LIE
ALGEBRAS AND BRAIDED GROUPS}
\\ \baselineskip 13pt{\ } {\ }\\ S. Majid\footnote{Royal Society
University Research Fellow and Fellow of Pembroke College,
Cambridge}\\ {\ }\\ Department of Applied Mathematics \& Theoretical
Physics\\ University of Cambridge, Cambridge CB3 9EW, U.K.
\end{center}

\begin{center}
December, 1993\end{center}
\vspace{10pt}
\begin{quote}\baselineskip 13pt
\noindent{\bf ABSTRACT}
We obtain an R-matrix or matrix representation of the Artin braid group acting
in a canonical way on the vector space of every (super)-Lie algebra or
braided-Lie algebra. The same result applies for every (super)-Hopf algebra or
braided-Hopf algebra. We recover some known representations such as those
associated to racks. We also obtain new representations such as  a non-trivial
one on the ring $k[x]$ of polynomials in one variable, regarded as a
braided-line. Representations of the extended Artin braid group for braids in
the complement of $S^1$ are also obtained by the same method.
\end{quote}
\baselineskip 22pt

\section{Introduction}
In this paper we apply some constructions from the theory of braided groups and
braided geometry\cite{Ma:introm} to obtain a new construction for matrix
solutions of the celebrated Quantum Yang-Baxter Equations (QYBE). Equivalently,
we provide a new and canonical class of representations of the Artin braid
group. The importance of such representations or R-matrices has been very
clearly established in the last few years and is one of the primary motivations
behind the celebrated quantum groups $U_q(g)$\cite{Dri}\cite{Jim:dif}.
Representations lead ultimately to link invariants and families of
representations to 3-manifold invariants.

By contrast to the theory of quantum groups, our construction is based on what
we believe to be a more primitive object, called a {\em braided Lie
algebra}\cite{Ma:lie}. The famous quantum groups $U_q(g)$ have
finite-dimensional braided-Lie algebras associated to them and one can work
with them instead of the quantum group. In this case our canonical braiding
reproduces the braiding associated to the quantum double\cite{Dri}.

More generally however, our notion of braided Lie algebra also includes as a
special case the notion of a rack, see e.g\cite{FenRou:rac}. In this case we
recover the rack braiding. Super-Lie algebras, super-racks and other much more
esoteric objects are also included in the theory.

The axioms of a braided-Lie algebra are recalled in the Preliminaries. They are
Lie algebra-like objects living in a braided tensor category with braiding
$\Psi$. Ordinary Lie algebras and ordinary racks are defined with $\Psi$ given
by the usual transposition. Their super-versions are defined with $\Psi=\pm 1$
according to a $\Z_2$-grading.
Unlike previous attempts to go beyond supersymmetry, we need not assume that
$\Psi^2=\id$.

In Section~3 we give a parallel theorem for Hopf algebras, super-Hopf algebras
and more generally, for braided-Hopf algebras\cite{Ma:bg}. The latter are Hopf
algebras living in our braided tensor category with background braiding $\Psi$.
This theory is more general because not every braided-Hopf algebra is the
enveloping algebra of a braided-Lie algebra, but in the case that it is, we
recover the results of Section~2. The example of a finite group and its
canonical braiding fits comfortably into either setting. We will also give some
more novel examples in Section~4, including one based on the anyonic line where
$\Psi$ is given by a root of unity. Many other important algebras of interest
in the theory of $q$-deformations are not naturally Hopf algebras but rather
braided ones.

Finally, we show in the Appendix how the extended Artin braid relations for
braids in the complement of the unknot in $S^3$ can be represented equally well
using the same techniques. We assume that we are given a cocommutative
representation of a braided-Hopf algebra in the sense that it is central in the
braided representation ring of the braided-Hopf algebra\cite{Ma:tra}.
Representations of such extended braid relations have been used by
knot theorists in \cite{LamPrz:hom} and elsewhere. In short, the techniques
which we use here are of quite wide applicability and this appendix
demonstrates one more instance of them.

Although we will not go as far as constructing knot and three-manifold
invariants from our canonical braiding, knot theory  nevertheless  enters in a
fundamental way. This is because we will be working throughout in a background
braided category with braiding $\Psi$. Usually one uses quantum groups etc to
construct such a braided category and hence to obtain knot-invariants: we
proceed in exactly the reverse direction by assuming that $\Psi$ is given and
doing all our proofs by drawing braids and tangles. These techniques and the
formulation of a large number of geometrical constructions of planes, lines,
matrices, groups, differential operators etc., is the topic of braided geometry
as developed over 30-40 papers by the author in the last few years. We refer to
\cite{Ma:introp}\cite{Ma:introm} for reviews and to
\cite{Ma:bg}\cite{Ma:tra}\cite{Ma:bos}\cite{Ma:skl}\cite{Ma:lin}\cite{Ma:fre}
for some of the basic theory.

\subsection*{Preliminaries} Here we recall very briefly the definition of
braided or quasitensor categories and the diagrammatic notation for them.
Firstly, a monoidal category consists of a category $\CC$ equipped with a
functor
$\tens:\CC\times \CC \to \CC$ and functorial isomorphisms
$\Phi_{V,W,Z}:V\tens (W\tens Z)\to
(V\tens W)\tens Z$ for all objects $V,W,Z$, and a unit
object $\und 1$ with
functorial isomorphisms $l_V:V\to \und 1\tens V,r_V:V\to
V\tens \und 1$ for all objects $V$.
The $\Phi$ should obey a well-known pentagon coherence
identity while the $l$ and $r$ obey triangle identities of
compatibility with $\Phi$\cite{Mac:cat}. We assume such a monoidal category and
suppress writing
$\Phi,l,r$ explicitly. A monoidal category also has an opposite tensor product
$\tens^{\rm op}:\CC\times\CC\to \CC$ defined in the obvious way.

A braided monoidal or quasitensor category $(\CC,\Psi)$ is a monoidal category
$\CC$ equipped further with a natural transformation $\Psi:\tens^{\rm op}\to
\tens$ called the {\em braiding} or quasisymmetry and subject to two hexagon
coherence identities. Explicitly, this means a collection of functorial
isomorphisms
$\Psi_{V,W}:V\tens W\to W\tens V$ for any two objects and such that
\eqn{psi-hex}{ \Psi_{V,W\tens Z}=\Psi_{V,Z}\circ\Psi_{V,W},\quad\Psi_{V\tens
W,Z}=\Psi_{V,Z}\circ\Psi_{W,Z}.}
One can deduce also that $\Psi_{V,\und 1}=\id=\Psi_{\und 1,V}$ for all $V$. If
$\Psi^2=\id$ then one of the
hexagons is superfluous and we have an ordinary symmetric monoidal category or
tensor
category. Braided monoidal categories were formally introduced in
\cite{JoyStr:bra}, while being known also to specialists in the representation
theory of quantum groups\cite[Sec. 7]{Ma:qua}.

Crucial for us is the following diagrammatic notation for working with
algebraic objects in braided categories. Firstly, we write all morphisms
pointing downwards (say) and in the case of the braiding morphism, we use the
shorthand
\eqn{Psi-bra}{\epsfbox{psinoarrow.eps}}
This distinguishes between $\Psi$ and $\Psi^{-1}$, while the hexagons
(\ref{psi-hex}) appear as
\eqn{Psi-hex-bra}{\epsfbox{smallhex.eps}}
The doubled lines refer to the composite objects $V\tens W$ and $W\tens Z$ in a
convenient extension of the notation. The coherence theorem for braided
categories says then that if two series of morphisms
built from $\Psi,\Phi$ correspond to the same braid then they compose to the
same morphism. The proof is just the same as Mac Lane's proof in the symmetric
case with the action of the symmetric group replaced by that of the Artin braid
group.

Finally, we take this notation further by writing any other
morphisms as nodes on a string connecting the inputs down to the outputs.
Functoriality of the braiding then says that morphisms $\phi:V\to Z$, $W\to Z$,
etc.  can
be pulled through braid crossings,
\eqn{Psi-funct}{\epsfbox{funcnode.eps}}
Similarly for $\Psi^{-1}$ with inverse braid crossings. The simplest example is
with $\CC={\rm SuperVec}$, the category of $\Z_2$-graded vector spaces and
braiding
\eqn{supertran}{\Psi(v\tens w)=(-1)^{|v|
|w|}w\tens v}
where $|\ |$ denotes the degree of a homogeneous element. Of course, this
example is not truly braided since $\Psi^2=\id$.

We recall also the celebrated Yang-Baxter equations or Artin braid relations.
Thus, a Yang-Baxter operator is a morphism $\check{R }:V\tens V\to
V\tens V$ such that
\eqn{YBop}{{\check{R}}_{23}\circ {\check{R}}_{12}\circ {\check{R
}}_{23}={\check{R }}_{12}\circ {\check{R }}_{23}
\circ {\check{R }}_{12}}
where the suffices refer to the copy of $V$ in $V\tens V\tens V$. If $V$ is an
ordinary vector space and everything is linear then we can write $\check{R
}=PR$ where $P:V\tens V\to V\tens V$ is the permutation operator. Then the
corresponding equation for $R$ is
\eqn{QYBE}{ R_{12}R_{13}R_{23}=R_{23}R_{13}R_{12}}
which is the so-called {\em quantum Yang-Baxter equation} (QYBE). The matrices
of such operators are called in physics `R-matrices'.

There is a close relation between R-matrices and braided categories for which
the objects are built on vector spaces. Obviously, if $\Psi$ is a braiding then
$\Psi_{V,V}$ is an invertible Yang-Baxter operator and hence when $V$ is a
finite-dimensional vector space we have an associated invertible R-matrix.
Conversely, any invertible R-matrix defines a braiding on the monoidal category
generated by $V$.

Note that the general theory of Sections~2,3 works in any braided monoidal
category. In this case we use the word `operator' etc here a bit loosely. On
the other hand, our examples in Section~4 are in a $k$-linear setting where $k$
is a field, and then our operators are indeed linear maps.

\section{Canonical braiding of a braided-Lie algebra}

We have introduced in \cite{Ma:lie} the notion of a braided-Lie algebra or
Lie-algebra-like object in a braided monoidal category $(\CC,\Psi)$ as
$(\CL,\Delta,\eps,[\ ,\ ])$ where $(\CL,\Delta,\eps)$ is a coalgebra in the
category and $[\ , \ ]:\CL\tens\CL\to \CL\tens\CL$ is the {\em braided Lie
bracket} and is required to obey
\ceqn{Lie}{ \epsfbox{Lie1.eps}\quad\qquad\epsfbox{Lie2.eps}\\
\epsfbox{Lie3.eps}}
We use here the diagrammatic notation described in the preliminaries. A
coalgebra in the category is defined in just the same way as an algebra, but
with arrows reversed. Thus, $\Delta:\CL\to \CL\tens \CL$, the comultiplication,
is coassociative in an obvious sense and $\eps:\CL\to \und 1$ is a counit for
it in the obvious sense. Explicitly,
\eqn{Liecoalg}{\epsfbox{Liecoalg.eps}}

 The condition (L1) is called the {\em braided-Jacobi} identity axiom, (L2) the
{\em braided-cocommutativity} axiom and (L3) the {\em coalgebra-compatibility
axiom}. We refer to \cite{Ma:lie} for the justification and full explanation of
these axioms. Suffice it to say that in a truly braided category the naive
notions of $\Psi$-anticommutativity and $\Psi$-Jacobi identity are not
appropriate and one needs a genuinely new idea. The new idea in \cite{Ma:lie}
is to allow ourselves a more general coalgebra $\Delta$ instead of the
primitive coalgebra structure $\Delta\xi=\xi\tens 1+1\tens\xi$ on $k\oplus g$
which is implicitly assumed in the theory of Lie algebras.

The basic theory of braided-Lie algebras has also been developed in
\cite{Ma:lie}. This includes such things as (in the Abelian category case) a
braided-enveloping bialgebra $U(\CL)$, braided-Killing forms and
braided-Casimirs etc. To this theory we want to add now the following theorem
announced in \cite{Ma:mex}.

\begin{theorem} Let $(\CL,\Delta,\eps,[\ ,\ ])$ be a braided-Lie algebra. Then
\[ \check{\bf R }=\ {{}\atop \epsfbox{Liebraid.eps}}=([\ ,\
]\tens\id)\circ(\id\tens\Psi)\circ(\Delta\tens\id):\CL\tens\CL\to \CL\tens
\CL\]
is a Yang-Baxter operator.
\end{theorem}
\begin{figure}
\[\epsfbox{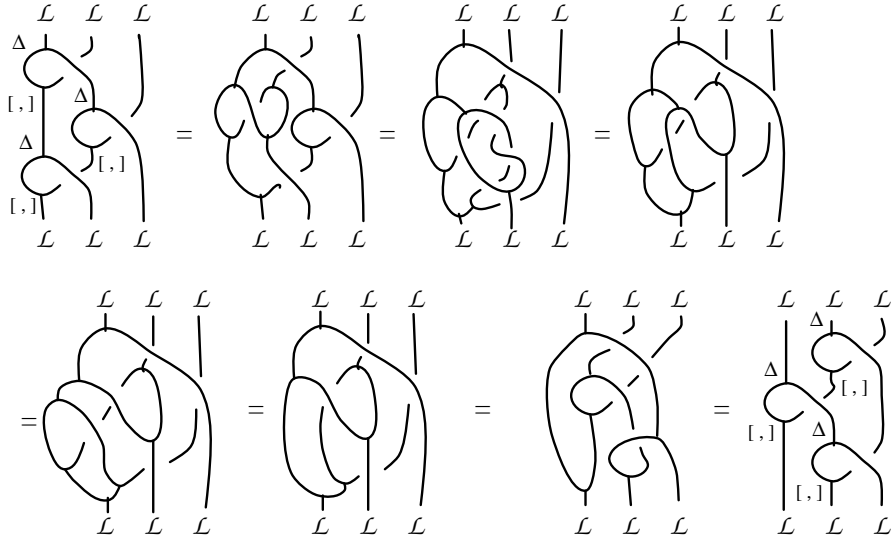}\]
\caption{Proof of Theorem~2.1}
\end{figure}
\proof We do this diagrammatically in Figure~1, using the notation explained in
the preliminaries. The vertices are $\Delta=\epsfbox{deltafrag.eps}$ and $[\ ,\
]=\epsfbox{prodfrag.eps}$ throughout.
The first expression is the right-hand side of (\ref{YBop}) for $\check{\bf R
}$ as stated. The first equality is (L3). The second equality is
coassociativity (\ref{Liecoalg}) and functoriality to put the diagram in a form
suitable for (L2), which is the third equality. The fourth equality is
coassociativity (\ref{Liecoalg}) again. The fifth then uses our braided-Jacobi
identity axiom (L1). The sixth is coassociativity once more and finally we use
functoriality to slide the diagram into the final form, which is the left hand
side of (\ref{YBop}) for $\check{\bf R }$. \endproof

Moreover, it is evident from its diagrammatic definition in \cite{Ma:lie} that
the braided enveloping algebra $U(\CL)$ is generated by $1$ and
$\CL$ with the relations
\eqn{bracom}{ \cdot\circ\check{\bf R }=\cdot}
of braided commutativity.

\section{Canonical braiding of a braided-Hopf algebra}

In this section, we further generalise the result of the last section to
associate to any braided-Hopf algebra
at all a canonical Yang-Baxter operator. Braided-Hopf algebras were introduced
by the author in \cite{Ma:bg}\cite{Ma:exa}\cite{Ma:eul}\cite{Ma:bra} as a
generalisation to braided categories of the usual notion of
Hopf algebra or super-Hopf algebra. Briefly, a braided-Hopf algebra means
$(B,\Delta,\eps,S)$ where firstly $B$ is a unital algebra in a braided monoidal
category. This means it comes equipped with product and unit morphisms $B\tens
B\to B$ and $\und1\to B$ respectively, obeying the obvious axioms of
associativity and unity. Secondly, $\Delta:B\to B\und\tens B$ and
$\eps:B\to\und 1$ form a coalgebra as already encountered in Section~2. We
require further that $\Delta$ is an algebra homomorphism where $B\und\tens B$
is the braided tensor product algebra (as also introduced by the author). This
forms a braided-bialgebra or bialgebra in a braided category. Finally, we
require an antipode $S:B\to B$ obeying axioms similar to those for quantum
groups or Hopf algebras, but as a morphism in our category. In the diagrammatic
notation with
$\Delta=\epsfbox{deltafrag.eps}$ and $\cdot=\epsfbox{prodfrag.eps}$, our axioms
read
\ceqn{hopfax}{ \epsfbox{bialg-ax.eps}\\ \epsfbox{ant-ax.eps}}

The general theory of braided-Hopf algebras has also been developed by now in a
diagrammatic form\cite{Ma:tra}\cite{Ma:bos}\cite{Ma:introp}. One has left and
right dual Hopf algebras (when $B$ has a dual object), regular actions and
coactions, braided-adjoint actions and coactions, cross products etc. We use
the braided-adjoint action
now.

\begin{theorem} Let $(B,\Delta,\eps,S)$ be a braided-Hopf algebra and $\Ad$ the
braided-adjoint action. Then
\[ \check{\bf R }=\ {{}\atop
\epsfbox{hopfbraid.eps}}
=(\Ad\tens\id)\circ(\id\tens\Psi)\circ(\Delta\tens\id):B\tens B\to B\tens B\]
is a Yang-Baxter operator.
\end{theorem}
\begin{figure}
\[\epsfbox{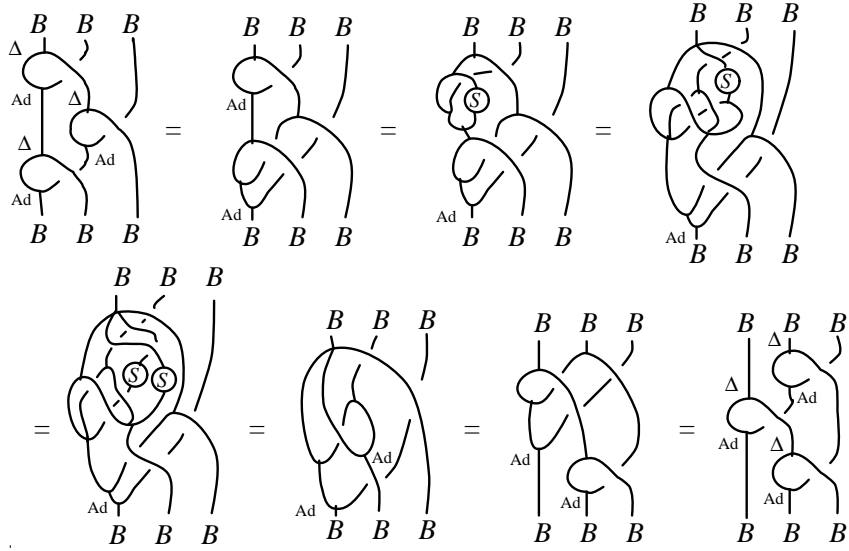}\]
\caption{Proof of Theorem~3.1}
\end{figure}
\proof This is given in diagrammatic form in Figure~2. Some of the
$\epsfbox{prodfrag.eps}$ vertices are the braided-adjoint action
$\Ad$\cite{Ma:exa}\cite{Ma:lie} and the rest are the product in $B$. The
$\epsfbox{deltafrag.eps}$ are the coproduct throughout. The first expression is
the right hand side of (\ref{YBop}) for $\check{\bf R }$ as stated. The first
equality
uses that $\Ad$ is indeed an action of $B$ on $B$. The second equality
substitutes the form of $\Ad$ in terms of the braided-Hopf algebra structure,
as shown in the definition of $\check{\bf R }$.
The third equality uses the bialgebra axiom that $\Delta$ is an algebra
homomorphism to the braided tensor product as on the left in (\ref{hopfax}). We
also adopt the convention that repeated applications of $\Delta$ can be
represented by multiple branches. Likewise for multiple products. This
convention expresses coassociativity and associativity respectively.  The
fourth equality is the lemma proven in \cite{Ma:tra} that $S$ is a
braided-anti-algebra homomorphism in the sense
$S\circ\cdot=\cdot\circ\Psi\circ(S\tens S)$. The fifth equality recognises a
loop involving the antipode and cancels it according to the left-hand antipode
axiom shown in (\ref{hopfax}). We also recognise the remaining antipode $S$ as
part of an application of $\Ad$. The sixth equality is coassociativity and
functoriality to push this $\Ad$ down to the bottom of the expression. Finally,
we use again that $\Ad$ is an action to obtain the left hand side of
(\ref{YBop}) for $\check{\bf R }$. \endproof

Moreover, it is obvious from coassociativity, associativity and the axioms for
the antipode that $B$ itself is braided-commutative in the sense
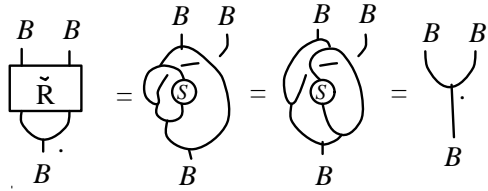
\[ \epsfbox{hbracom.eps}\]

Let us note also that the axioms of a braided-Hopf algebra in (\ref{hopfax})
are symmetric under the operations of left-right reflection and braid crossing
reversal, up-down reflection and braid crossing reversal, and rotation by 180
degrees. As explained in \cite{Ma:introm}, it means that the diagrammatic
method always gives three theorems for the price of one. Applying these
symmetries to Figure~2 and its associated lemmas gives
\[\check{\bf R }=\ {{}\atop \epsfbox{hbravar3.eps}}\qqquad \check{\bf R }=\
{{}\atop \epsfbox{hbravar1.eps}}\qqquad \check{\bf R }=\ {{}\atop
\epsfbox{hbravar2.eps}}\]
as three other Yang-Baxter operators. $B$ is also braided-commutative with
respect to the first and braided-cocommutative with respect to the second and
third.

\section{Examples}

In this section we describe various examples and special cases of the general
constructions above. Throughout this
section we work over a field $k$ of characteristic zero. The same results apply
more generally with appropriate care. As far as I know, only the case in
subsection~4.4 was known before from the theory of racks and its generalisation
in subsection~4.5 from Drinfeld's quantum double construction for Hopf algebras
and later from \cite{Wor:sol}. The rest appear to be a product of our
construction, as announced recently in the conference proceedings\cite{Ma:mex}.

\subsection{Ordinary Lie Algebras}

Note that an ordinary Lie algebra obeys these axioms if one
puts $[1,\xi]=\xi$, $[\xi,1]=0$, $[1,1]=1$ and
\eqn{ordLie}{\CL=k\oplus g,\quad \Delta 1=1\tens 1,\ \eps 1=1,\quad
\Delta\xi=\xi\tens
1+1\tens\xi,\ \eps\xi=0,\quad\forall \xi\in g.}
So this structure $\Delta,\eps$ is implicit for an ordinary Lie algebra but we
never think about it because it has this standard form. This was our motivation
in
\cite{Ma:lie}.

\begin{propos} Let $V=k\oplus g$ and define the linear map
\cmath{ \check{\bf R }(1\tens 1)=1\tens 1,\quad {\check{\bf R }} (1\tens
\xi)=\xi\tens 1,\quad
{\check{\bf R }} (\xi\tens 1)=1\tens\xi\\
{\check{\bf R }} (\xi\tens\eta)=\eta\tens\xi+[\xi,\eta]\tens
1,\qquad\qquad\quad \forall
\xi,\eta\in g.}
Then ${\check{\bf R }} $ is a braiding {\em iff} $[\ ,\ ]:g\tens g\to g$ obeys
the Jacobi
identity. It has minimal polynomial
\eqn{minpoly}{({\check{\bf R }} ^2-\id)({\check{\bf R }} +\id)=0}
 {\em iff} $[\ ,\ ]$ is non-zero and antisymmetric.
\end{propos}
\proof The forward direction is a special case of Theorem~2.1 where we view
$\CL=k\oplus g$ as a braided-Lie algebra with trivial braiding $\Psi$ as
explained above. On the other hand, at least in this setting, one can compute
it explicitly and see that it is reversible. Thus
\align{{\check{\bf R }}_{23}\circ {\check{\bf R }}_{12}\circ {\check{\bf R
}}_{23}(\xi\tens\eta\tens\zeta)&&\equad
=\zeta\tens\eta\tens\xi+\zeta\tens[\xi,\eta]\tens 1+[\xi,\zeta]\tens\eta\tens
1\\
&&\quad +[\eta,\zeta]\tens 1\tens\xi+[\xi,[\eta,\zeta]]\tens 1\tens 1\\
{\check{\bf R }}_{12}\circ {\check{\bf R }}_{23}
\circ {\check{\bf R }}_{12}(\xi\tens\eta\tens\zeta)&&\equad
=\zeta\tens\eta\tens \xi+[\xi,\zeta]\tens\eta\tens 1+[\eta,\zeta]\tens
1\tens\xi\\
&&\quad +[\eta,[\xi,\zeta]]\tens 1\tens 1 +\zeta\tens [\xi,\eta]\tens
1+[[\xi,\eta],\zeta]\tens 1\tens 1}
so that the only condition is the Jacobi identity in the form that $[\xi,\ ]$
acts like a Lie derivation. The braid relations involving the basis element $1$
are all empty. Secondly, we compute
\[ (\check{\bf R }^2-\id)\circ (\check{\bf R
}+\id)(\xi\tens\eta)=([\xi,\eta]+[\eta,\xi])\tens 1+1\tens
([\xi,\eta]+[\eta,\xi])\]
so this vanishes {\em iff} the bracket is antisymmetric. Finally, we compute
\cmath{ (\check{\bf R }\pm\id)(\xi\tens\eta)=\eta\tens\xi+[\xi,\eta]\tens 1\pm
\xi\tens \eta\\
(\check{\bf R }^2-\id)(\xi\tens\eta)=[\eta,\xi]\tens 1+1\tens[\xi,\eta]\\
(\check{\bf R }+\id)^2(\xi\tens\eta)=2\xi\tens\eta+[\eta,\xi]\tens
1+1\tens[\xi,\eta]+2\eta\tens\xi+2[\xi,\eta]\tens 1}
which are all non-zero for some $\xi,\eta$ if $[\ ,\ ]$ is non-zero. Hence in
this case (\ref{minpoly}) is the minimum polynomial. Conversely, if this is the
minimum-polynomial then
in each case there exist $\xi,\eta$ such that the expression is non-zero. In
particular, $(\check{\bf R }^2-\id)\ne 0$ implies that $[\xi,\eta ]\ne 0$ for
some $\xi,\eta$. \endproof.

This says that the definition of a Lie algebra
is mathematically completely equivalent to looking for a braiding of a certain
form. We say accordingly that a Yang-Baxter operator obeying (\ref{minpoly}) is
of {\em Lie type}.

We next discuss the braided-enveloping algebra. To avoid confusion here we
denote the basis element of $k$ in $k\oplus g$ by
$\lambda$ rather than $1$ as above. So our starting point is that that $\Psi$
trivial (usual transposition) and
\cmath{ \Delta \lambda=\lambda\tens\lambda, \quad \Delta
\xi=\xi\tens\lambda+\lambda\tens\xi,\quad \eps\lambda=1,\quad \eps\xi=0\\
{}[\lambda,\lambda]=\lambda,\quad [\lambda,\xi]=\xi,\quad [\xi,\lambda]=0,\quad
\forall\xi\in g}
extends a usual Lie algebra $g$ to a braided Lie algebra $\CL=k\oplus g$.
The braided enveloping algebra in this case is $U(\CL)=\widetilde{U(g)}$, a
homogenised form of the enveloping algebra $U(g)$. It is an ordinary bialgebra
since the braiding $\Psi$ is trivial. On the other hand we know from the
definition of $U(\CL)$ that it is braided-commutative in the sense
(\ref{bracom}) relative to $\check{\bf R }$. Moreover, everything descends to
the quotient $\lambda=1$ so we recover the usual enveloping algebra too as
braided-commutative in this sense, a fact which is in any case evident from the
form of $\Psi$.

Note also that if we allow $\lambda^{-1}$ then this form of $\widetilde{U(g)}$
has $\xi\lambda^{-1}$ primitive and $\lambda$ group-like, and forms a Hopf
algebra. It is only for this part of its structure that antisymmetry of the Lie
bracket of $g$ is needed. For a general braided-Lie algebra I do not know any
very natural notion of antisymmetry for the bracket.

Finally, we give the concrete matrix version in which we choose a basis
$g=\{x_i\}$ for $i=1,2,\cdots ,n-1$ and $k=\{x_0\}$ for $V=k\oplus g$. Then
\eqn{g-R}{{\bf R}=\pmatrix{1&0&0&0\cr 0&I&0 & c\cr 0&0&I&0\cr 0&0&0&I}}
where $I$ are identity matrices and $c^i{}_{jk}$ are the structure constants of
the bracket $[\ ,\ ]$ on $g$. The basis for $V\tens V$ used here is $\{x_0\tens
x_0,x_0\tens
x_j,x_i\tens x_0,x_i\tens x_j\}$. Explicitly,
\[ {\bf R}^0{}_i{}^k{}_j=c^k{}_{ij},\ {\bf
R}^i{}_j{}^k{}_l=\delta^i{}_j\delta^k{}_l,\  {\bf
R}^0{}_0{}^i{}_j=\delta^i{}_j={\bf R}^i{}_j{}^0{}_0,\ {\bf R}^0{}_0{}^0{}_0=1\]
and zero for the rest. This obeys the QYBE and has minimal polynomial of Lie
type {\em iff} $c$ defines a non-zero Lie algebra. The quantum R-plane or
Zamolodchikov algebra
\[ x_ix_j=x_bx_a{\bf R}^a{}_i{}^b{}_j\]
for this R-matrix recovers the homogenised enveloping algebra above in our
basis. The indices here range $0,\cdots,n-1$ and summation of the repeated
indices is understood.

Finally, we note that homogenised Lie algebras have recently been studied in
\cite{LeBSmi:hom}\cite{LeBBer:spa} as examples of a kind of non-commutative
geometry based on `projective line modules'. It would be interesting to try
connect this with the braided-geometrical picture developed above.

\subsection{Super-Lie Algebras}

A super-Lie algebra is a $\Z_2$-graded or `super' vector space $g$ with a
degree-preserving map $[\ ,\ ]:g\tens g\to g$ obeying the axiom of
graded-antisymmetry and the graded-Jacobi identity:
\[ [\xi,\eta]=(-1)^{|\xi||\eta|+1}[\xi,\eta],\quad
[[\xi,\eta],\zeta]+[\eta,[\xi,\zeta]](-1)^{|\xi||\eta|}=[\xi,[\eta,\zeta]]\]
on homogeneous elements $\xi,\eta,\zeta\in g$. One can view any super-Lie
algebra as a braided-Lie algebra $\CL=k\oplus g$ in the category of super
vector spaces with braiding given by super-transposition (\ref{supertran}) and
the remaining structure
as in subsection~4.1 in (\ref{ordLie}). The $k$ part of $\CL$ is given degree
zero.

\begin{propos} Let $g$ be a $\Z_2$-graded vector space and $V=k\oplus g$ with
$k$ given degree zero, and $[\ ,\ ]:g\tens g\to g$ a degree-preserving linear
map. Then
\cmath{ \check{\bf R }(1\tens 1)=1\tens 1,\quad {\check{\bf R }} (1\tens
\xi)=\xi\tens 1,\quad
{\check{\bf R }} (\xi\tens 1)=1\tens\xi\\
{\check{\bf R }} (\xi\tens\eta)=(-1)^{|\xi||\eta|}\eta\tens\xi+[\xi,\eta]\tens
1,\qquad\qquad\quad \forall
\xi,\eta\in g.}
obeys the braid relations {\em iff} $[\ ,\ ]:g\tens g\to g$ obeys the
graded-Jacobi
identity. Moreover, it has minimal polynomial (\ref{minpoly}) {\em iff} $[\ ,\
]$ is graded-antisymmetric and non-zero.
\end{propos}
\proof That $\check{\bf R }$ obeys the braid relations follows from Theorem~2.1
in the category of $\Z_2$-graded vector spaces where $V$ is viewed as a
braided-Lie algebra in this category as explained. Conversely, an explicit
computation along the same lines as the proof of Proposition~4.1 gives that the
braid relations force $[\ ,\ ]$ to obey the graded Jacobi identity. Similarly
for the minimal polynomial by explicit computation. \endproof

The braided-enveloping algebra $U(\CL)$ in this case is a homogenised
super-bialgebra version of the enveloping super-Hopf algebra $U(g)$. It is the
Zamolodchikov or quantum plane algebra for the matrix ${\bf R}$ corresponding
to $\check{\bf R }$ in this case.

This generalisation of the preceding subsection is immediate because the
category is not truly braided, i.e. one has $\Psi^2=\id$ and hence all the
properties familiar in the category of vector spaces. The same applies if we
work in any symmetric monoidal category of vector spaces with $\Psi^2=\id$ and
the same form of coproduct $\Delta$ and $[\ ,\ ]$ on an object $\CL=k\tens g$.
The analogue of Proposition~4.2 recovers the obvious axioms of a $\Psi$-Lie
algebra as studied, for example, in \cite{Gur:yan}. We still find
(\ref{minpoly}) for $\check{\bf R}$ even for this more general case.

\subsection{Matrix braided-Lie algebras}

The data we need is a matrix solution  $R\in M_n\tens M_n$ of the QYBE
which is bi-invertible. The `second inverse' $\widetilde R$ which we suppose
here is
characterised by
\[\widetilde{R}^i{}_a{}^b{}_l
R^a{}_j{}^k{}_b=\delta^i{}_j\delta^k{}_l
=R^i{}_a{}^b{}_l\widetilde{R}^a{}_j{}^k{}_b.\]
We assume summation of repeated indices throughout this section. These
$R,\widetilde{R}$ generate a braided monoidal category $\CC$ and this has an
associated braided group $\Aut(\CC)$\cite{Ma:bra}\cite{Ma:bg} which has in
turn, a braided-Lie algebra $\CL$. Explicitly\cite{Ma:skl}\cite{Ma:lie},
\eqn{L(R)-coalg}{\CL=k^{n^2}=\{u^i{}_j\},\quad \Delta u^i{}_j=u^i{}_k\tens
u^k{}_j,\quad \eps
u^i{}_j=\delta^i{}_j.}
\eqn{B(R)-stat}{ \Psi(u_J\tens u_L)= u_K\tens u_I \Psi^I{}_J{}^K{}_L;\quad
\Psi^I{}_J{}^K{}_L=R^{j_0}{}_a{}^d{}_{k_0}
R^{-1}{}^a{}_{i_0}{}^{k_1}{}_b
R^{i_1}{}_c{}^b{}_{l_1} {\widetilde R}^c{}_{j_1}{}^{l_0}{}_d}
\eqn{L(R)}{{}[u_I,u_J]=u_K
c^K{}_{IJ};\qquad c^K{}_{IJ}=\widetilde{R}^{a}{}_{i_1}{}^{j_0}{}_b
R^{-1}{}^b{}_{k_0}{}^{i_0}{}_c R^{k_1}{}_e{}^c{}_d R^d{}_{a}{}^{e}{}_{j_1} }
where we write $I=(i_0,i_1)$ etc as multi-indices. We changed conventions here
from \cite{Ma:lie} to lower indices for the $\{u_I\}$. There is also a nice
compact notation used in physics where
subscripts refer to the positions in a matrix tensor product (as in the QYBE
above). In this notation,
\cmath{ \Delta \vecu=\vecu\tens\vecu,\ \eps\vecu=\id\\
\Psi(R_{12}^{-1}\vecu_1\tens R_{12}\vecu_2)=\vecu_2 R_{12}^{-1}\vecu_1
R_{12},\quad R_{21}[\vecu_1,R_{12}\vecu_2]=\vecu_2 R_{21}R_{12}.}

\begin{corol} Let $R\in M_n\tens M_n$ be a bi-invertible solution of the QYBE
and $\CL(R)$ its associated matrix braided-Lie algebra. Then the associated
canonical braiding from Theorem~2.1 is
\[ {\check{\bf R }} (u_J\tens u_L)=  u_K\tens u_I {\bf R}^I{}_J{}^K{}_L;\quad
{\bf R}^I{}_J{}^K{}_L=R^{-1}{}^{d}{}_{k_0}{}^{j_0}{}_{a}
R^{k_1}{}_{b}{}^{a}{}_{i_0}R^{i_1}{}_c{}^b{}_{l_1} {\widetilde
R}^c{}_{j_1}{}^{l_0}{}_d\]
and ${\bf R}\in M_{n^2}\tens M_{n^2}$ necessarily obeys the QYBE.
\end{corol}
\proof We compute the canonical braiding for the matrix braided-Lie algebra
above. In fact, the necessary computation was done already in the proof of
\cite[Prop. 5.2]{Ma:lie} in the course of computing the relations of $U(\CL)$.
We need only the matrix form of $\Delta$ and the formulae for
$\Psi=\epsfbox{braid.eps},[\ ,\ ]=\epsfbox{prodfrag.eps}$ in
(\ref{L(R)-coalg})--(\ref{L(R)}).  Hence from Theorem~2.1 we conclude that
$\check{\bf R }$ obeys the QYBE too. \endproof

The braided enveloping algebra $U(\CL)$ for this class was computed and
identified in \cite[Prop. 5.2]{Ma:lie} as the braided-bialgebra of $B(R)$ of
`braided matrices' as introduced in \cite{Ma:exa}. This is the associative
algebra generated by $1$ and $\vecu=\{u^i{}_j\}$
with the braided-commutativity relations (\ref{bracom}) which are now
\eqn{B(R)}{u_Ju_L= u_Ku_I {\bf R}^I{}_J{}^K{}_L,\quad {\rm i.e.}\quad
R_{21}\vecu_1R_{12}\vecu_2= \vecu_2 R_{21} \vecu_1 R_{12}}
where the second puts two of the $R$'s to the left and uses the matrix
notation. The coproduct $\Delta \vecu=\vecu\tens \vecu$ extends as an algebra
homomorphism $B(R)\to B(R)\und\tens B(R)$ to the braided tensor product algebra
 determined by $\Psi$, i.e. according to the axiom on the left in
(\ref{hopfax}). Note that the motivation in \cite{Ma:exa} for $B(R)$ was as a
braided-version of quantum or super matrices, with braid statistics $\Psi$,
i.e. the generators are to be regarded as, by definition, the
braided-commutative ring of co-ordinate functions
on a braided space. Hence it is remarkable that this $B(R)$ is also the
enveloping algebra of a braided-Lie algebra. We obtain in the corollary a new
and conceptual proof that the matrix $\bf R$ that describes its relations
indeed obeys the QYBE.

This class of examples generalises those of Section~4.1 for ordinary Lie
algebras and Section~4.2 for super-Lie algebras, as well as including the case
of Lie algebras defined in an obvious way relative to any background braiding
where $\Psi^2=\id$. The way to obtain these from the notion of braided-Lie
algebras is explained in \cite{Ma:lie}. One uses $\bar\chi={\vecu-\id\over
\hbar}$
along with $1$ as generators of $U(\CL)$ in place of $1$ and $\vecu$, where $R$
is parametrised in such a way that $R_{21}R=O(\hbar)$. The standard R-matrices
associated to semisimple Lie algebras $g$ in \cite{FRT:lie} then give
deformations as braided-bialgebras of their homogenised enveloping algebras
$\widetilde U(g)$ from this point if view.

Unfortunately, the simplest non-trivial example of a matrix braided-Lie algebra
has to be four-dimensional. We mention the standard one from \cite{Ma:lie},
namely the braided-Lie algebra $\CL=gl_{2,q}$ with basis
\[ gl_{2,q}=\{\pmatrix{a& b\cr c& d}\},\quad t=q^{-1}a+qd,\quad x={b+c\over
2},\quad y={b-c\over 2i},\quad z=d-a\]
if we work over $\C$. The braided-Lie bracket is obtained from (\ref{L(R)})
with the standard $sl_2$ R-matrix, and given explicitly in \cite[Example
5.5]{Ma:lie}. The braided-enveloping algebra here in terms of the $\chi$
variables is a deformation of $U(gl_2)$ or from another point of view, of
$\widetilde{U(sl_2)}$.

The canonical braiding $\check{\bf R }$ from Theorem~2.1 for this example is
the braided-commutativity relations for the algebra of $2\times 2$ braided
matrices\cite{Ma:exa} and with the braided-determinant
\[ {\rm BDET}\pmatrix{a& b\cr c& d}={q^2\over(q^2+1)^2}t^2-q^2x^2-q^2 y^2-
{(q^4+1)q^2\over 2(q^2+1)^2}z^2+\left({q^2-1\over q^2+1}\right)^2{q\over 2}
tz\]
it can be viewed as a braided $q$-deformation of the algebra of functions on
Minkowski space with its Lorentzian metric\cite{Ma:mec}\cite{Mey:new}. These
$t,x,y,z$ are the non-commutative spacetime co-ordinates. The algebra here also
agrees with the proposal for $q$-Minkowski space based on spinors in the
approach \cite{CWSSW:lor}\cite{OSWZ:def}.
We note that the FRT bialgebra $A({\bf R})$ associated to this canonical
braiding has a Hopf algebra quotient $SO_q(1,3)$, the $q$-Lorentz group in the
interpretation above. On the other hand, this is also closely related to the
dual of the quantum double of $U_q(su_2)$. These points are described in detail
elsewhere.

The braided-enveloping algebra here of $2\times 2$ braided matrices is also
isomorphic to a degenerate form of the 4-dimensional Sklyanin algebra as shown
in \cite{Ma:skl}, so the latter has the R-matrix form (\ref{B(R)}). More
recently, some remarkable homological properties of braided-matrix algebras
have been found in \cite{LeB:hom}.

\subsection{Finite Groups and Racks}

A {\em rack} is a set $X$ and a map $X\times X\to X$ denoted $x\times y\mapsto
{}^xy$ obeying the `rack-identity'
\[ {}^{({}^{\scriptstyle x}y)}({}^x z)={}^x({}^yz),\qquad \forall x,y,z\in X.\]
One usually adds to this that the map ${}^x(\ )$ is bijective for each $x$, but
we do insist on this here. One may also have conventions in which the notation
is $y^x$ rather than ${}^x y$. Such objects have a long history and some
applications in algebraic topology\cite{FenRou:rac}. It is easy to see that
every rack provides an example of a braided-Lie algebra if we take as our
category $\CC$ as the category of sets, with tensor product provided by the
direct product of sets, and with the usual permutation map as $\Psi$. This is a
symmetric monoidal category rather than a truly braided one. We just take
\[ [x,y]={}^xy,\quad \Delta x=x\times x\]
and note that the axiom (L1) in (\ref{Lie}) then becomes the rack identity
above, while the others are empty. At the level of sets the braiding from
Theorem~2.1 is
\[ \check{\bf R }(x\times y)=[x,y]\times x\]
and recovers the braiding associated to a rack in \cite{FenRou:rac}.

For a $k$-linear setting over a field we let $\CL=kX$, the vector space with
basis $X$, and the above definitions extended $k$-linearly. So $\CL$ is the
coalgebra with basis $X$ and all basis elements grouplike. Then we have a
(trivially braided) braided-Lie algebra in the category of vector spaces and
the canonical braiding from Theorem~2.1 is just
\[ \check{\bf R }(x\tens y)=[x,y]\tens x.\]
The braided-enveloping algebra $U(\CL)$ consists of the algebra generated by
elements of $X$ modulo the relations $xy=[x,y]x$ for all $x,y$. This is the
bialgebra generated by the rack monoid. Here the rack monoid is the free monoid
generated by symbols from $x,y$ modulo such relations, and is also a classical
construction for racks.

Our examples of matrix braided-Lie algebras in subsection~4.3 are a deformation
of a mixture of some Lie-algebra like elements as in subsection~4.1 and some
rack-like elements. In the $gl_{2,q}$ example mentioned there, the rack-like
element is proportional to the `time' direction $t$ and the Lie-algebra like
elements are the `space' directions $x,y,z$.

We see that even when the braiding is trivial, the notion of a braided-Lie
algebra in (\ref{Lie}) is still useful. Moreover, it is more general than a
rack because we are free to specify a more general coproduct $X\to X\times X$
than the diagonal map, as long as we obey (L1)--(L3) in the category of sets.
The simplest way to obey (L2) is for $\Delta$ to be cocommutative.

The classic example of a rack is a group with the rack operation
$[x,y]=xyx^{-1}$. The braiding $\check{\bf R }$ in the $k$-linear setting is
then the braiding associated to the quantum double Hopf algebra
$D(X)$\cite{Dri}. The latter is defined for $X$ finite but the rack point of
view is more general and works for any group. The associated 3-manifold
invariants in this case are well-known, see for example \cite{FreYet:bra}. On
the other hand, we have super-racks, etc. just as easily as examples of
braided-Lie algebras, and their associated braidings may prove more
interesting.

\subsection{Ordinary Hopf Algebras}

If $H$ is any Hopf algebra then it acts on itself by the Hopf-algebra adjoint
action $\Ad_h(g)=\sum h\o g S h\t$ where we use the notation $\Delta h=\sum
h\o\tens h\t$ of \cite{Swe:hop} for the coproduct. Theorem~3.1 reduces for
ordinary Hopf algebras (with trivial braiding) to
\eqn{brahopf}{{\check{\bf R }} (h\tens g)=\sum \Ad_{h\o}(g)\tens
h\t,\qquad\forall h,g\in H.}
One can easily verify in a couple of lines that $\check{\bf R }$ obeys the
braid relations. This was perhaps first explicitly remarked in \cite{Wor:sol}.
The case of $H$ a group algebra clearly reduces us to the rack braiding in
subsection~4.4.

Once again, this braiding can be viewed as originating in Drinfeld's quantum
double construction $D(H)$\cite{Dri}, this time applied to a general
finite-dimensional Hopf algebra $H$. Drinfeld introduced $D(H)$ as a
quasitriangular Hopf algebra defined by generators and relations in a basis.
Here the quasitriangular structure $\CR\in D(H)\tens D(H)$ obeys Drinfeld's
axioms which are such as to ensure that its image in any representation obeys
the QYBE. We introduced a form of this in \cite{Ma:phy} built explicitly on the
vector space $H^*\tens H$ with product
\[ (a\tens h)(b\tens g)=\sum \<Sh\o,b\o\>b\t a\tens h\t g \<
h\th,b\th\>,\qquad\forall a,b\in H^*,\ h,g\in H\]
and tensor product unit and coalgebra. In writing this we switch also to
conventions with $H$ and $H^{*\rm op}$ (the opposite algebra) as sub-hopf
algebras, rather than Drinfeld's original conventions with $H,H^{*\rm cop}$
(the opposite coalgebra). Let $\{e_a\}$ be a basis of $H$ and $\{f^a\}$ a dual
basis then
\[ \CR=\sum_a(f^a\tens 1)\tens (1\tens e_a)\]
is Drinfeld's quasitriangular structure in these conventions.

\begin{propos} $D(H)$ acts on $H$ by
\[  (1\tens h)\la g=\sum h\o g Sh\t,\quad (a\tens 1)\la g=\sum \<a, h\o\>h\t\]
and the associated braiding is (\ref{brahopf}).
\end{propos}
\proof It is easy to see that this defines an action of $D(H)$ (this is
modelled on quantum mechanics and could be called the `Schroedinger
representation' of the quantum double). Then the action of $\CR$ is $\CR\la
(h\tens g)=\sum_a (f^a\tens 1)\la h\tens (1\tens e_a)\la g=\sum h\t\tens
(1\tens h\o)\la g=\sum h\t\tens \Ad_{h\o}(g)$ giving exactly (\ref{brahopf})
for the corresponding $\check{\bf R }$. \endproof

Thus the braiding from Theorem~3.1 does not give anything genuinely new for an
ordinary Hopf algebra. It is slightly more general than the braiding coming
from the quantum double in that it does not require $H$ to be finite
dimensional, but this issue too can be dealt with in other ways\cite{Dri}. On
the other hand, it is still a useful observation, as is the fact which is
obvious from (\ref{brahopf}) that $\cdot\circ \check{\bf R }=\cdot$ holds in
$H$. See for example \cite{AndDev:ext} where such an observation recently
proved very useful for some Hopf algebraic constructions.

\subsection{Super Hopf Algebras}

To obtain something new from Theorem~3.1 we can consider Hopf algebras in
categories other than the usual one of vector spaces. The simplest setting is
that of $\Z_2$-graded or super Hopf algebras. These are defined in the obvious
way with all maps degree-preserving, where the degree of a term in a tensor
product is the sum of the degrees in a homogeneous decomposition. Plenty of
super-Hopf algebras are known, not least in algebraic
topology\cite{MilMor:str}.

The braiding in this case is
\[ \check{\bf R }(h\tens g)=\sum  \Ad_{h\o}(g)\tens (-1)^{|h\t||g|} h\t;\quad
\Ad_{h}(g)=\sum h\o g Sh\t (-1)^{|h\t||g|}\]
where it is assumed that all tensor product elements are decomposed
homogeneously.

\subsection{The Braided Line $k[x]$}

The previous subsection is still not a truly braided example of Theorem~3.1.
Truly  braided-Hopf algebras were first introduced and studied by the author
through a number of papers. In this subsection we compute Theorem~3.1 for the
simplest of these\cite{Ma:any}, where the braiding is still a factor but not
necessarily $\pm 1$ as it was in subsection~4.6.

The braided-line $k[x]$ as an algebra is nothing other than the polynomials in
one variable. However, we regard it as an algebra in the category of
$\Z$-graded vector spaces. As such, it has a braiding
\eqn{anybraid}{ \Psi(x^m\tens x^n)=q^{nm}x^n\tens x^m}
where $q\ne 0,1$ is a fixed but otherwise arbitrary element of $k$. The ideas
here are from \cite{Ma:any}. We define $\Delta x=x\tens 1+1\tens x$, $\eps x=0$
and extend to products as a braided-Hopf algebra according to (\ref{hopfax}).
It is easy to see that
\eqn{anydelta}{ \Delta x^m=\sum_{r=0}^m [{m\atop r};q] x^r\tens x^{m-r},\quad
[{m\atop r};q]={[m;q]!\over [r;q]![m-r;q]!},\quad [m;q]={q^m-1\over q-1}.}
The $q$-integers and $q$-binomial coefficients here are
well-known\cite{And:ser} but we use them in a novel way as defining a
braided-Hopf algebra structure\cite{Ma:fre}. Using the lemma that $S$ is a
braided-antialgebra map we have also that
\eqn{anyS}{ S x^m=q^{m(m-1)\over 2}(-x)^m}
to complete the braided-Hopf algebra structure. Finally, we note that this
braided-Hopf algebra approach to $q$-analysis derives the standard
$q$-derivative
\eqn{anydif}{  (\del_q f)(x)={f(qx)-f(x)\over x(q-1)},\quad \del_q
x^m=[m;q]x^{m-1}}
as infinitesimal translations\cite{Ma:fre}, a point of view which generalises
at once to $n$-dimensional quantum plane algebras.

\begin{lemma} The braided adjoint action of $k[x]$ on itself as in Theorem~3.1
is
\[ \Ad_f(g)(x)=f(x^2(1-q)\del_q)g(x),\qquad \forall f,g\in k[x]\]
\end{lemma}
\proof We know from the general theory of braided-Hopf algebras\cite{Ma:introp}
that the braided-adjoint action is an action and also that it acts on itself as
a braided module-algebra. The latter condition means in the present case that
it acts as a braided-derivation
\[ \Ad_x(fg)=\Ad_x(f)g+\cdot\circ \Psi(\Ad_x\tens
f)g=\Ad_x(f)g+L_q(f)\Ad_x(g);\quad L_q(f)(x)=f(qx).\]
Since $\Ad_x(x)=xx-qxx=(1-q)x^2$ we deduce from this $q$-derivation property of
the adjoint action that $\Ad_x(x^n)=x^{n+1}(1-q^n)=x^2(1-q)\del_q x^n$. Since
$\Ad_x$ is an action, we deduce the result stated. Explicitly, this has on
monomials the form
\[ \Ad_{x^m}(x^n)=(1-q^n)\cdots (1-q^{n+m-1})x^{n+m}=(1-q)^m{[n+m-1;q]!\over
[n-1;q]!}x^{n+m}.\]
\endproof

We note in passing that our derivation here depends strongly on the properties
of $\Ad$ proven in \cite{Ma:lin}\cite{Ma:exa} using the same novel diagrammatic
techniques as in Figure~2. If we try to compute it directly from
(\ref{anydelta})--(\ref{anyS}) on monomials then we derive the novel
$q$-identity
\[ \sum_{r=0}^m [{m\atop r};q] q^{r(r-1)\over 2} (-1)^r
q^{rn}=(1-q)^m{[n+m-1;q]!\over [n-1;q]!}.\]
This is the content of the lemma from the point of view of $q$-analysis.

\begin{corol} The braiding $\check{\bf R }:k[x]\tens k[x]\to k[x]\tens k[x]$
obtained from Theorem~3.1 is
\[ \check{\bf R }(x^m\tens x^n)=\sum_{r=0}^m [{m\atop r};q] q^{n(m-r)}(1-q)^r
{[n+r-1;q]!\over [n-1;q]!} x^{n+r}\tens x^{m-r}.\]
\end{corol}
\proof We compute from Theorem~3.1 in our case. Thus from the formula above for
$\Delta$ we have
\[ \check{\bf R }(x^m\tens x^n)=\sum_{r=0}^m [{m\atop r};q] \Ad_{x^r}(x^n)\tens
x^{m-r} q^{n(m-r)}\]
and putting in the form of $\Ad$ computed in Lemma~4.5 gives the result stated.
\endproof

We now provide a braided-geometrical picture of this $\check{\bf R }$ as an
operator on polynomials in two variables. Thus we distinguish the two copies of
$k[x]$ in both the input and output, so that $\check{\bf R }:k[y]\tens k[x]\to
k[x]\tens k[y]$ say. Next we consider these variables to be non-commuting with
the quantum-plane relations, i.e.
\[ \check{\bf R }:k[x,y;q]\to k[x,y;q],\quad k[x,y;q]={k\<x,y\>\over yx-qxy}.\]
This is a purely notational device because this algebra has a basis
$\{y^nx^m\}$ so as a linear space can be identified with $k[y]\tens k[x]$, but
also has a basis $\{x^my^n\}$ and so can be identified with the linear space
$k[x]\tens k[y]$ as well.

\begin{propos} $\check{\bf R }$ in the form $k[x,y;q]\to k[x,y;q]$ is the
operator
\[ \check{\bf R }=e_q^{x^2(1-q)\del_{q,x}|\del_{q,y}}=\sum_{m=0}^\infty
{(x^2(1-q)\del_{q,x})^m\del_{q,y}^m\over [m;q]!}\]
where the $|$ denotes that the $q$-exponential is to be understood as ordered
in the form shown and $\del_{q,x}$ and $\del_{q,y}$ are as in (\ref{anydif})
acting on $x,y$ respectively.
\end{propos}
\proof Note that the expression is a well-defined operator since on any
polynomial the power-series always terminates. The $q$-exponential is the
standard one except that we have adopted the ordering convention stated. This
is such that we have
\eqn{anyTay}{ e_q^{x\del_{q,y}}f(y)=(\Delta f)(x,y)=f(x+y)}
where we recall that $\Delta$ is an algebra homomorphism to the braided tensor
product algebra which we identify as $k[x]\und\tens k[y]=k[x,y;q]$ for the
braiding (\ref{anybraid}). This (\ref{anyTay}) describes  a `braided-Taylors
theorem' as explained in \cite{Ma:fre}, where it is also generalised to
$n$-dimensions. Applying the braided-adjoint representation
$x\mapsto\Ad_x=(1-q)x^2\del_{q,x}$ to both sides of (\ref{anyTay}) allows us to
recompute $\check{\bf R }$ from Theorem~3.1 as
\[ \check{\bf R
}(f(y)g(x))=f(x^2(1-q)\del_{q,x}+y)g(x)
=e_q^{x^2(1-q)\del_{q,x}|\del_{q,y}}f(y)g(x)\]
which is the form stated. Here $f,g$ are arbitrary polynomials and a general
polynomial in $x,y$ can be written as a linear combination of such products.
\endproof

In this form, our Yang-Baxter operator $\check{\bf R }$ has some similarities
with the quasitriangular structure $\CR$ of $U_q(sl_2)$. To see this we note
that classically one can embed $sl_2$ inside the Witt algebra of `vector
fields' on $k[x]$ by
\[ L_0=x{d\over dx},\quad L_1=x^2{d\over dx},\quad L_{-1}={d\over dx}\]
so $\check{\bf R }$ resembles the factor $e_q^{(1-q)L_{1}\tens L_{-1}}$
ocurring in the formula for $\CR$ in \cite{KirRes:rep}, cf\cite{Dri} and
elsewhere. On the other hand, there is no Gaussian factor $q^{L_0\tens L_0}$ as
to be found there.

If one works  over $\C[[\hbar]]$ rather than over a field and sets
$q=e^{\hbar}$ then $\check{\bf R }$ has a an expansion
\[ \check{\bf R }=P\circ(\id+\hbar r+O(\hbar^2));\quad r=x{d\over dx}\tens
x{d\over dx}-{d\over dx}\tens x^2{d\over dx}\]
which $r$ is necessarily an operator realisation $k[x]\tens k[x]\to k[x]\tens
k[x]$ of the Classical Yang-Baxter equation (CYBE). Here $P$ is permutation. To
obtain this formula one can work from Corollary~4.6 or else from
Proposition~4.7 provided one remembers the contribution (the first term in $r$)
coming from the fact that the output of $\check{\bf R }$ has to be viewed in
$k[x]\tens k[y]$ while its input is viewed in $k[y]\tens k[x]$. In this case
$r$ is indeed the image in the Witt algebra of the Drinfeld-Jimbo
solution\cite{Dri}\cite{Jim:dif} of the CYBE on $sl_2$ when represented as
vector fields on $k[x]$. The connection with theory of quantum groups is
provided by the `bosonisation theorem' introduced in \cite{Ma:bos}. In the
present case this turns constructions on $k[x]$ into equivalent ones on the
Hopf algebra $U_q(b_+)$ in \cite{Dri}.

On the other hand, we have obtained this $\check{\bf R }$ starting from nothing
other than $k[x]$ regarded as a braided-Hopf algebra (the braided-line) and
Theorem~3.1. Moreover, we can suppose that $q\in k^*$ is a root of unity and
proceed with the same calculations. More precisely, we can repeat the above
calculations for the braided-Hopf algebra $U_n(k)=k[x]/x^n$ introduced in
\cite{Ma:any}, where $q$ is a primitive $n$'th root of $1$. This `anyonic line'
braided-Hopf algebra is $n$-dimensional and hence $\check{\bf R }$ corresponds
to a matrix solution of the QYBE in $M_n\tens M_n$. An elementary computation
gives for example,

\begin{example} The braiding from Theorem~3.1 applied to the anyonic line for
$n=3$ has minimal polynomial
\[ (\check{\bf R }^2-1)^2(\check{\bf R }-q)=0\]
and corresponding matrix solution of the QYBE
\[ {\bf R}=\pmatrix{1&0&0 &0&0&0   &0&0&0\cr
                    0&1&0 &0&0&0   &0&0&0\cr
                    0&0&1 &0&1-q&0 &0&0&0\cr
                    0&0&0 &1&0&0   &0&0&0\cr
                    0&0&0 &0&q&0   &0&0&0\cr
                    0&0&0 &0&0&q^2 &0&$q$-1&0\cr
                    0&0&0 &0&0&0   &1&0&0\cr
                    0&0&0 &0&0&0   &0&q^2&0\cr
                    0&0&0 &0&0&0   &0&0&q}\]
where $q^3=1$ is a primitive root and the basis is $\{1\tens 1,1\tens
x,\cdots,x^2\tens x^2\}$.
\end{example}

This braided-line which we have studied above is only the very simplest example
of a braided-Hopf algebra. The next simplest is probably the quantum plane
$k[x,y;q]$ this time regarded itself as a braided-Hopf algebra with braiding
provided by the standard $sl_2$ R-matrix corresponding to the Jones knot
polynomial\cite{Ma:poi}. Explicitly,
\cmath{\Psi(x\tens x)=q^2 x\tens x,\  \Psi(x\tens y)=q y\tens x,\  \Psi(y\tens
y)=q^2 y\tens y\\
 \Psi(y\tens x)=q x\tens y+(q^2-1)y\tens x.}
Using this, one obtains the braided adjoint action on generators as
\[ \Ad_x(x)=x^2(1-q^2),\ \Ad_x(y)=(1-q^2)xy,\ \Ad_y(x)=yx(1-q^2),\
\Ad_x(y)=y^2(1-q^2).\]
Extending this to higher products along the lines above, one obtains the
braided-vector fields for the braided-adjoint action as
\[ \Ad_x=(1-q^2)x(x\del_{q,x}+y\del_{q,y}),\quad
\Ad_y=(1-q^2)y(x\del_{q,x}+y\del_{q,y})\]
where $\del_{q,x}$ and $\del_{q,y}$ are the partial derivatives on the quantum
plane\cite{WesZum:cov} in the form obtained by an infinitesimal translation in
\cite{Ma:fre}.
{}From this one can compute the braiding $\check{\bf R }$ from Theorem~3.1 as
represented by $q$-deformed vector fields on a quantum plane. We  have not
found an explicit exponential formula for it along the lines of
Proposition~4.7.

This class of examples generalises further to any quantum plane algebra of the
R-matrix type. The necessary braided-Hopf algebra structure, braided
differential calculus and R-binomial theorem are in \cite{Ma:fre}. For example,
the $q$-Minkowski space example with generators $t,x,y,z$ fits into this
setting with additive braided-Hopf algebra structure found in \cite{Mey:new}.
In these cases the appropriate $q$-exponential map needed for the analogue of
Proposition~4.7 is not yet known.
In a different direction, we can take the free braided-Hopf algebra
$k\<x_1,\cdots,x_n\>$ with braiding determined by $R$ as a generalisation of
the example $k[x]$ above. In this case the relevant exponential map is provided
in \cite{Ma:fre}.

\appendix

\section{Extended braid relations and braided cocommutativity}

Here we give an application of the notion, introduced by the author in
\cite{Ma:bra} of
a  braided-cocommutative module of a braided-bialgebra. We do not need a
braided-antipode in this section. Given $B$ a bialgebra in a braided tensor
category $\CC$ as in Section~3, a $B$-module in the category means
$(V,\alpha_V)$ where $V$ is an object and $\alpha_V:B\tens V\to V$ is a
morphism obeying the obvious notion of an action of $B$ on $V$. We say that $B$
is {\em braided-cocommutative} with respect to $V$ (or simply that the module
is {\em braided cocommutative} when the Hopf algebra is understood) if
\eqn{V-cocom}{ \epsfbox{V-cocom.eps}}
So for example, the braided-enveloping bialgebra $U(\CL)$ in Section~2 acts
cocommutatively on $\CL$ by $[\ , \ ]$. The braided-Hopf algebras that arise by
transmutation from quantum groups\cite{Ma:bra} are likewise
braided-cocommutative with respect to all the corresponding braided-modules
that come from modules of the quantum group. So this is a large class.

Now, for any $B$-module $(V,\alpha_V)$ we define the associated operator
\[ U_{B,V}=\ {{}\atop\epsfbox{Ualpha.eps}}\
=(\id\tens\alpha_V)\circ\Delta:B\tens V\to B\tens V\]
Working with this is equivalent to working with our original module since we
can recover the latter as $(\eps\tens\id)\circ U_{B,V}=\alpha_V$. We have

\begin{theorem} Let $(V,\alpha_V)$, $(W,\alpha_W)$ be modules of a
braided-bialgebra $B$, with $(W,\alpha_W)$ cocommutative. Then the
corresponding operators $U$
obeys the extended braid relations
\[ U_{B,V}\circ\Psi_{W,V}\circ U_{B,W} \circ \Psi_{V,W}=\Psi_{W,V}\circ U_{B,W}
\circ \Psi_{V,W}\circ U_{B,V}\]
\end{theorem}
\begin{figure}
\[ \epsfbox{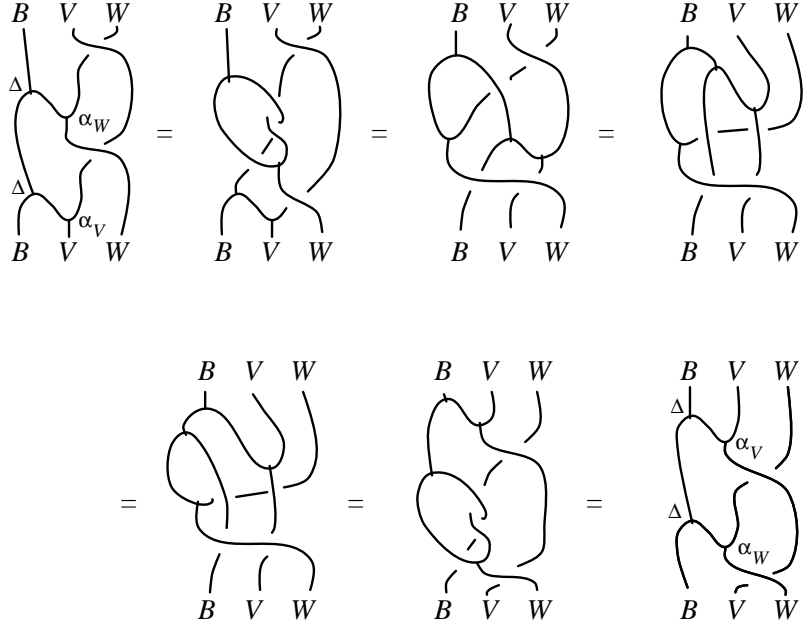}\]
\caption{Proof of Theorem~A.1}
\end{figure}
\proof This is shown in our diagrammatic notation in Figure~3. The
$\epsfbox{prodfrag.eps}$ vertices are the module action $\alpha_V$ or
$\alpha_W$ while the $\epsfbox{deltafrag.eps}$ vertices are the coproduct
$\Delta$ throughout. We begin with the left hand side of the extended braid
relations with $U$ of the form stated above and $\Psi=\epsfbox{braid.eps}$ as
usual. The first equality is (\ref{V-cocom}) for $W$. The second slides this
group containing $\alpha_W$ over to the left using functoriality. The third
likewise pushes the group containing $\alpha_V$ up to the top of the
expression. The fourth is coassociativity of $\Delta$. The fifth pulls the
$\alpha_W$ vertex down and to the right. The sixth now uses the cocommutativity
condition (\ref{V-cocom}) again for $W$, to obtain the right hand side of the
extended braid relations as required.
\endproof

We note that this proof used nothing more than coassociativity of the coproduct
$\Delta$ and the cocommutativity axiom (\ref{V-cocom}) as it was introduced in
\cite{Ma:bra}, being essentially equivalent to it when we demand it for all
$V$. Another equivalent way to write the condition is that $\alpha_{V\tens
W}\isom \alpha_{W\tens V}$ by the braiding $\Psi_{V,W}$, i.e. that the module
$V$ is central in the category of $B$-modules, up to the trivial isomorphism
provided by the background braiding. This is the key property of
representations of ordinary groups (which are commutative up to ordinary
transposition) and was the main motivation behind the theory of braided groups
in \cite{Ma:tra}. See \cite[Eqn. (64)]{Ma:introp} for the connection with
physics from this point of view.

To describe a class of examples we observe that exactly the same definitions
and theorem hold for a braided-commutative module of a braided-Lie algebra
$\CL$ (as defined in \cite{Ma:lie}) so it is not necessary to work with
an entire braided-bialgebra here. In particular, $\CL$ acts on itself by
$\alpha_{\CL}=[\ ,\ ]$, the adjoint representation of the braided-Lie algebra.
So we have
\[ U_{12}\circ \Psi_{23}\circ U_{12}\circ \Psi_{23}=\Psi_{23}\circ U_{12}\circ
\Psi_{23}\circ U_{12}\]
as operators on $\CL\tens\CL\tens\CL$. Such matrix representations of the
extended Artin braid group are useful in defining invariants of links in the
complement of the trivial knot,  as explored in \cite{LamPrz:hom}. The
canonical extension to an action of $U(\CL)$ puts us into the setting of
Theorem~A.1.

Note also that one can of course turn all these diagram-proofs upside-down.
Then we have another operator $U_{V,B}$ for every right $B$-comodule $V$. This
time the upside-down (\ref{V-cocom}) becomes the condition of $V$ a commutative
right-comodule with respect to $B$, as studied in\cite{Ma:bg}. An example is
provided by $B=B(R)$ as the bialgebra of braided matrices. It has a
braided-coaction on the Zamolodchikov or quantum R-plane algebra $V$ with
generators $\{x_i\}$ and coaction $\beta(x_i)=x_j\tens u^j{}_i$, which is
braided-commutative because it comes from transmutation\cite{Ma:lin}. Hence we
can apply Theorem~A.1 in dual form. This gives another point of view on the
braided-commutativity relations (\ref{B(R)}) of $B(R)$.

\end{document}